# Library and Information Science Research in Indian Universities: Growth, Core Journals, Keywords and Collaboration Patterns

**Dr. Swapan Kumar Patra**
Career Advancement Research Fellow
Industrial Engineering Department
Faculty of Engineering and the Built Environment
Tshwane University of Technology
Pretoria, Republic of South Africa

Email: skpatra@gmail.com; PatraSK@tut.ac.za

**Abstract**

This article maps Library and Information Science (LIS) research in Indian universities. As the two prominent citation databases, Web of Science and Scopus have very limited coverage of Indian LIS journals, the publications generated by the library and science departments of about 114 selected Indian universities and the two national institutions of importance in LIS research were extracted from *Library, Information Science & Technology Abstracts* (LISTA). The relevant publication records were analyzed using scientometrics and Social Network Analysis (SNA) tools. The study traces the growth of publications, prominent keywords, leading journals where the articles are published and the institutional collaboration patterns of Indian university publications. The results show that there is a growth in scholarly publications from Indian universities in LIS. However, the numbers of publications are limited to only a few universities and national institutes of importance. The maximum LIS research outputs are published in Indian journals. Bibliometrics related investigations are the most important research areas. Located in major cities of India, the productive institutes show healthy collaboration. The study concludes with some observations which may be useful for formulating policies in LIS research in India.

**Introduction**

Library and Information Science (LIS) research in India is about a century old and has rich research heritage. India was the pioneer among the South Asian countries in imparting LIS education and training. Although, the progress in LIS education and training in India was rather slow in the initial years, it is seen that during the last twenty years of the twentieth century there has been a significant growth in LIS publications. Thereafter the reforms introduced since the beginning of the 21st century and the all-pervasive role of Information and Communication Technology (ICT) have boosted the LIS research and training in India[1]. LIS education and research in India is largely carried out in the universities and a few national institutes. A comprehensive mapping of LIS research is required to understand the characteristics of the Indian LIS research including identifying prominent institutions, authors & publications, studying collaborations patterns and so on. Hence, this study is an attempt to map LIS research in Indian universities and national institutes.

**Literature Review**

The history of LIS education has been traced back to the initiatives by two students of Melville Dewey during the early twentieth century[2,3]. Among the South Asian countries LIS education first began in India. The earliest evidence of library training in India goes back to 1903 when a library staff member at the Central Hindu College in Bañaras (Varanasi) was sent to the Imperial (National) Library, Calcutta, for in-house training. In 1911, W. A. Borden, librarian of the Young Men's Institute, New Haven, Connecticut, USA started the first training program for library workers at the Central Library in Baroda, Gujarat[3]. In 1910, the then Maharaja of Baroda, encouraged the librarian from the United States to organize a public library system in the state of Baroda. However, the formal library school started in Panjab University, Lahore, with a course in library science in 1915. A. D. Dickinson established a training course in 1915 in Panjab University (now in Pakistan). This training course was considered as the second library school established in the world at that time. Thereafter, other universities and library associations also started setting up library schools and courses in India[3,4]. However, all these efforts in pre-independent India were short lived and lasted only for a couple of years[1]. Before independence, only five universities (Andhra, Banaras, Bombay, Calcutta and Madras) offered the diploma course in library science[5]. The real Library education and training in India in the field of LIS was recognized after India's independence in 1947. In independent India, the momentum to LIS research was given by Dr. S.R. Ranganathan (considered as the father of library science, in India) at the University of Delhi. His various initiatives in LIS research and education placed India on the global map. Under his visionary leadership, two new programs; a Master degree program in 1948 and a PhD in Library Science started in 1950. So, University of Delhi was the first institution in India that offered advanced degree programs in LIS. Moreover, University of Delhi was the only institute that offered master's degree until 1965. However, many other schools introduced Master of Library Science (M. Lib. Sci.) programs after 1965. This included Bañaras Hindu University, University of Bombay, Panjab University, and Aligarh Muslim University[1]. So, it can be said that the higher education in LIS in India, is about nine decades old, and is centered in universities. Over a period of time, LIS has grown and developed into a full-fledged discipline. Many advanced level courses are being imparted by university departments, institutions, library associations and specialized institutions. Presently about 120 universities are offering bachelor's degree, 78 are offering Post Graduate degree, 21 are offering two-year integrated course, 16 universities are offering M.Phil. degree, and 63 are offering Ph.D. degree[5]. Beside these, two national institutes of importance, the Documentation Research and Training Centre (DRTC) of Indian Statistical Institute in Bangalore, and Indian National Scientific Documentation Centre (INSDOC) in New Delhi focused on professionals training for special and industrial libraries & information centres. [INSDOC later known as National Institute of Science Communication and Information Resources (NISCAIR). In 2021, NISCAIR merged with National Institute of Science, Technology and Development Studies (NISTADS) and formed National Institute of Science Communication and Policy Research] The strong ICT components in the curricula were the hallmark of the courses offered by the two institutes. The more ICT related course structure made these two institutions different from other programs in their environment and products[6].

Along with the education and training, the genesis of LIS journals to publish the research output in LIS field in India is also now more than a century old. The first Indian LIS journal named *Library Miscellany* was published in 1912. However, the journal stopped in 1919. Thereafter, a number of LIS journals have been published from India. Although, many scholarly LIS journals published from India was short lived, some are still in existence. For example, *IASLIC Bulletin*, *Library Herald* and so on are now more than 50 years old and still in existence. The exceptional contribution of Dr. S. R. Ranganathan was not only in the establishment of LIS education and training but in research. He had also established a couple of journals which continue to be brought out till date[7].

With a century of librarianship behind it, India maintains its Third World leadership in research in library education and literature[6,8]. Indian government agencies, for example the University Grants

Commission (UGC) and Indian Council of Social Science Research (ICSSR) are playing important roles in promoting LIS research. These government institutes award fellowships to scholars to pursue their doctoral studies. Beside this Defense Scientific Information and Documentation Centre (DESIDOC), Delhi, also provides Junior Research Fellowship (JRF) in LIS[9].

There are a few studies that have attempted to map the LIS research in India. The source data of such studies have been from Library and Information Science Abstracts[10,11], and doctoral dissertations data[12, 13]. This study is an attempt to fill the gap of understanding the nature of LIS research in Indian universities.

**Research Questions**

In line with the previous studies, this paper deals with at following research questions. What is the literature growth patterns in Indian LIS research outputs from the universities over the years? Which are the productive universities and their publications in the preferred journals? What are the research areas? What are the institutional collaboration patterns? This study will address the collaboration trends, network structures and core groups at the institution levels.

**Methodology**

Due to the very little coverage of Indian journals by Web of Science and Scopus[14], alternative sources that exhaustively index Indian LIS journals are to be considered. Google Scholar is the next natural choice but owing the indexing inconsistencies and lack of user friendliness in downloading the requisite data, it was overlooked. So, Library and Information Science & Technology Abstracts (LISTA), one of the comprehensive abstracting databases that abstracts 560 core LIS journals including most Indian journals was considered for this study. Besides, the database also includes books, research reports and conference proceedings. The database was searched using different parameters focusing on Indian literature produced from universities till the year 2018.

Based on the data, the literature growth patterns are traced, core journals are identified, and social network analysis tools are used to visualize and measure the characteristics of the network such as density, clustering coefficient, components and mean distances. The centrality measures such as degree centrality, closeness centrality and betweenness centrality are used to identify the most prominent actor in the collaboration[15, 16, 17]. The opensource version of social networking software UCINET[18] and NetDraw[19] are used to draw and analyze the collaboration network of institutions and VoS viewer was used to draw keyword clustering maps[20].

**Results**

Records of about 114 universities are searched and downloaded. Among the list of 114 universities, autonomous colleges and two institutes of national importance, about 98 universities has some publications. About 16 universities or institutes do not have any publications listed in LISTA. So, this study is based on 3,160 records from 98 universities. After removing the duplicate (because of collaborative articles) 2,783 records are finally selected for this study. Among these, 2,425 items are articles (87 percent) Book Chapter 118(4.2 percent), Book Review 104 (3.7 percent) Case Study 46 (1.6 percent), Proceeding 28 (1 percent) and others 72 (2.5 percent). The records are further analyzed to trace the year wise growth pattern of literature, core journals, prominent keywords and collaboration patterns among Indian universities. The following sections deals with the results.

**Growth of literature**

Although the publications by Indian LIS researchers started as early as 1964 onwards, the contributions by the Indian universities were minimal (Figure 1). During the initial years, the two institutions of national importance were predominant in LIS publications. Documentation Research and Training Centre (DRTC) Indian Statistical Institute, Bangalore and Indian National Scientific Documentation Centre (INSDOC). Although the Indian Universities published in LIS field, the publication trends were sporadic and inconsistent. From the year 1964 to the year 1999 the publications by Indian Universities were negligible. This was perhaps because of the nature of LIS education being carried out in the country. During that period, the LIS education was being carried out mostly by the 'traditional way' and the late adoption of ICT in Indian libraries as well as education systems[21].

However, LIS education underwent metamorphic changes during the 1980s, both structurally and functionally. Indian universities adopted the University Grants Commission (UGC) recommendations based on Ranganathan Committee Report (1969), and the UGC Subject Panel on Report LIS in 1980s[21]. Universities started graduate, postgraduate, MPhil and PhD programs. Besides, Documentation Research and Training Centre, Bangalore, and Indian National Scientific Documentation Centre, New Delhi, started advanced training in library science with emphasis on documentation services. The 'information' component was given a high priority from late 1970s and the nomenclature of library science was changed to library and information science as stipulated by the UGC[21]. Also, the ICT Revolution had given a major boot in research in LIS field.

**Figure 1 Cumulative scholarly publication from Indian universities**

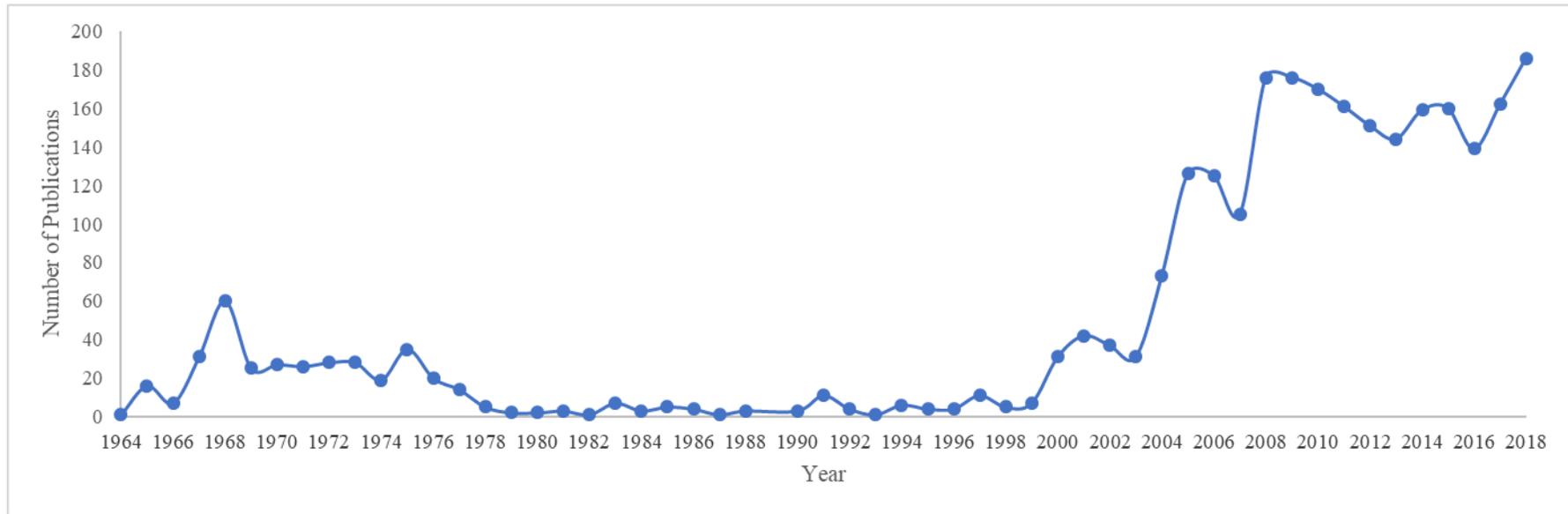

The period after the 1999 saw a major boot in LIS research in India. Initiation of automation activities and development of software packages in libraries started in 1990s[22]. Table 1 shows the productive universities or institutions in LIS research in India. DRTC and NISCAIR are the two most productive institutes in LIS research. This is because of these institutions specality in LIS education and research.

**Table 1 Top LIS schools and their publication counts**

| Sl No | Name of the University | State | Total publications till the year 2018 |
|---|---|---|---|
| 1. | Indian Statistical Institute, Bangalore / Documentation Research and Training Centre | Karnataka | 599 |
| 2. | University of Delhi | Delhi | 282 |
| 3. | National Institute of Science Communication & Information Resources | Delhi | 147 |
| 4. | University of Mysore | Karnataka | 139 |
| 5. | University of Kashmir | Jammu & Kashmir | 105 |
| 6. | Guru Nanak Dev University | Punjab | 94 |
| 7. | Aligarh Muslim University | UP | 76 |
| 8. | Indira Gandhi National Open University | Delhi | 75 |
| 9. | Banaras Hindu University | UP | 73 |
| 10. | Karnatak University | Karnataka | 70 |
| 11. | Panjab University | Punjab | 64 |
| 12. | University of Kerala | Kerala | 62 |
| 13. | Annamalai University | Tamil Nadu | 59 |
| 14. | Gulbarga University | Karnataka | 54 |
| 15. | Bangalore University | Karnataka | 52 |
| 16. | Kuvempu University | Karnataka | 51 |
| 17. | University of Calcutta | West Bengal | 49 |
| 18. | University of Madras | Tamil Nadu | 49 |
| 19. | University of Jammu | Jammu & Kashmir | 48 |
| 20. | Mangalore University | Karnataka | 41 |
| 21. | Sambalpur University | Orissa | 41 |
| 22. | University of Burdwan | West Bengal | 41 |
| 23. | Alagappa University | Tamil Nadu | 39 |
| 24. | Punjabi University | Punjab | 38 |
| 25. | University of Calicut | Kerala | 34 |
| 26. | Jadavpur University | West Bengal | 32 |
| 27. | Manonmaniam Sundaranar University | Tamil Nadu | 30 |
| 28. | Mizoram University | Mizoram | 29 |
| 29. | Pondicherry University | Tamil Nadu | 27 |
| 30. | University of Pune | Maharashtra | 27 |

Although, the overall LIS publication output from universities are growing, the individual university level publication the LIS research, it shows quite low performance. The two institutes of national importance (DRTC, NISCAIR) followed by the following universities, University of Mysore, University of Kashmir, University of Delhi have more than 100 publications during the past 54 years (1964-2018) rest others have few publications. The result shows that, unlike the other subject, research in LIS in universities are perhaps not storng and need to be strengthened.

Also the institutes located in the 'core' for example institutes located in the metropolitan cities like Delhi, Bangalore are more productive. The institutes located in the 'periphery' for example the LIS school from northeastern part of India is less productive. The more productivity is because of the better infrastructure and facilities of these institutes. This assumptions may be strengthen with further research.

**Core journals**

About 2,782 research items including research articles are published in 205 different outlets of publications including the journals, monograms, seminar series case and so on. SRELS Journal of Information Management published the maximum number of articles and ranked number one in terms of total publications from India. The top journals are which published about half of the total articles of Indian LIS output are shown in table 2. This is inconformity of earlier studies that the Indian LIS researchers prefer to publish in Indian journals[10, 23, 7].

**Table 2 the top journals of Indian LIS research output**

| Sl. No | Name of the Journal | Number of publications | Cumulative total |
|---|---|---|---|
| 1. | SRELS Journal of Information Management | 437 | 437 |
| 2. | DESIDOC Journal of Library & Information Technology | 271 | 708 |
| 3. | Library Philosophy & Practice | 264 | 972 |
| 4. | Annals of Library & Information Studies | 176 | 1148 |
| 5. | Information Studies | 79 | 1227 |
| 6. | Journal of Library & Information Science | 66 | 1293 |
| 7. | Herald of Library Science | 57 | 1350 |
| 8. | Scientometrics | 49 | 1399 |
| 9. | International Information & Library Review | 44 | 1443 |
| 10. | DESIDOC Bulletin of Information Technology | 30 | 1473 |
| 11. | IFLA Conference Proceedings | 29 | 1502 |

**Research trends using keywords analysis**

Intellectual structure of a given subject can be mapped by utilizing co-word analysis[24]. The status and trends of any subject can be mapped keywords extracted from the set of relevant journals.

Various approaches are used to map the trends in development of a subject for example the keyword used in the title, abstract, author assigned, keywords or the combinations of different types of assigned keywords[25,26].

For this study keywords are extract from the downloaded records. There are altogether 21,050 keywords assigned to the articles 2,424 articles. Rest other articles are not assigned any keywords. The figure 1 shows the cumulative keywords clustering. The clusters are further divided into two period. All records till the year 2000 are grouped in one cluster and records from 2001-2018 are in the second cluster. The selection of period is divided into two period because as per the figure 1 the significant growth of the literature was observed only after the year 2000. It may be correlated with the extensive use of ICT in Indian libraries and library services.

The keywords from the records were retrieved from the "Subject Term", "Thesaurus Terms" and "Author-Supplied Keywords". The keyword clustering maps were constructed accordingly. From the initial period (1964-2000) period there was only 460 records and 169 unique keywords. These keywords formed 34 clusters. The largest cluster consists of 12 keywords and bibliometrics is the most prominent keyword among them.

**Figure 2 A Keyword clustering map of LIS research in Indian universities during 1964-2000**

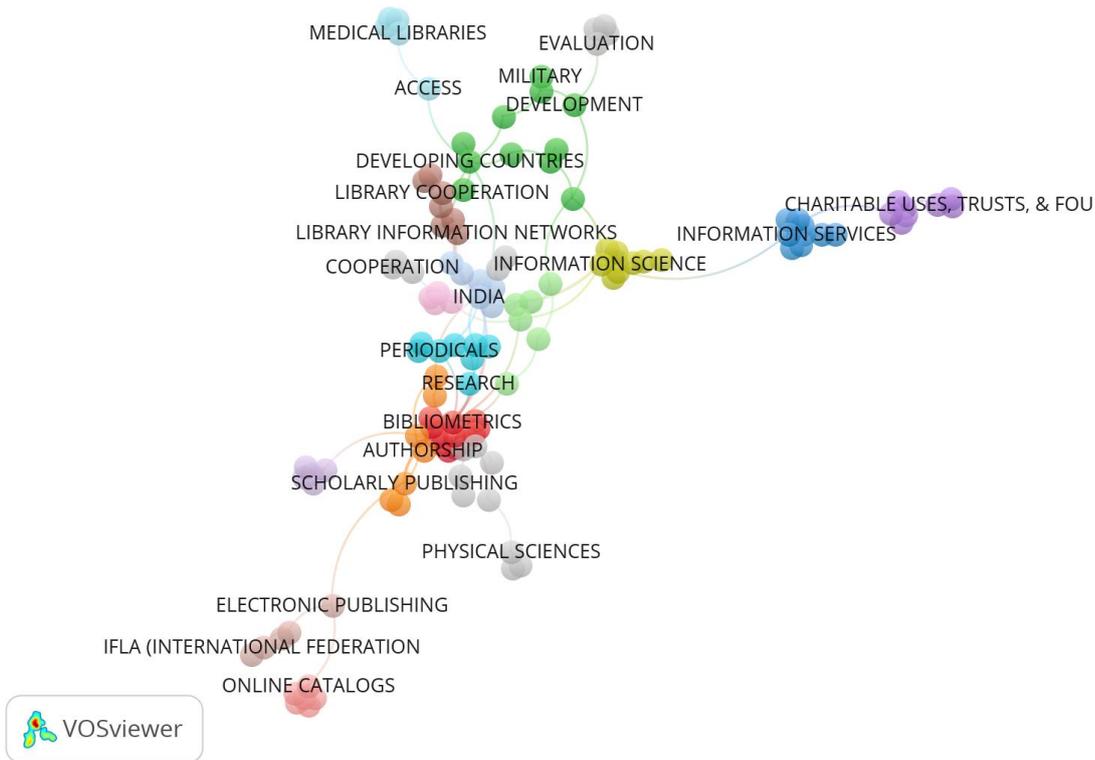

From the period 2001 to 2018 there are 2,323 records and 6857 keywords, formed 499 clusters. The largest component in cluster has 92 keywords.

The cluster of all keywords during the period 1964-2018 have 6922 keywords and formed 505 clusters. The largest cluster has 91 keywords. The top keywords during the different time period are shown in table 3

**Figure 2 B Keyword clustering map of LIS research in Indian universities during 2001-2018**

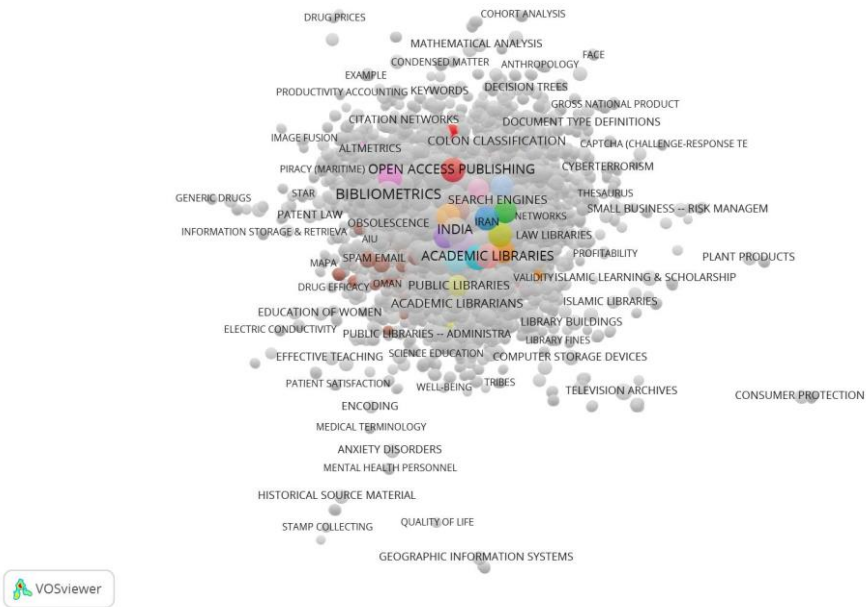

**Figure 2 C Keyword clustering map of LIS research in Indian universities during 1964-2018**

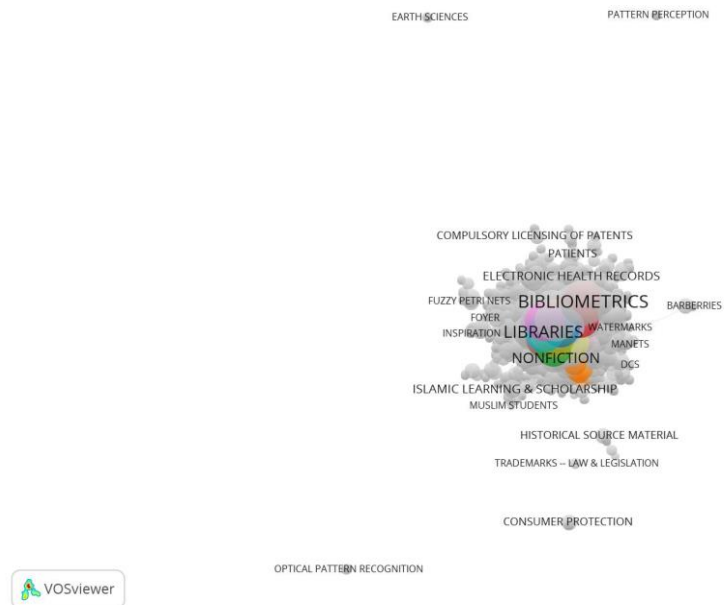

Table 3 shows the different keyword's occurrence in different span of time. The table shows that Bibliometrics is the most frequently occurring keywords in Indian University LIS research. This shows the research strength in bibliometrics studies.

**Table 3: Frequently occurring keywords during different span of time.**

|        | Period 1964-2000          |        | Period 2001-2018                        |        | Period 21964-2018                       |        |
|--------|---------------------------|--------|-----------------------------------------|--------|-----------------------------------------|--------|
| Sl. No | Keywords                  | Degree | Keywords                                | Degree | Keywords                                | Degree |
| 1.     | bibliometrics             | 25     | bibliometrics                           | 518    | bibliometrics                           | 539    |
| 2.     | India                     | 15     | academic libraries                      | 356    | academic libraries                      | 360    |
| 3.     | information science       | 11     | libraries                               | 338    | libraries                               | 344    |
| 4.     | information services      | 9      | India                                   | 285    | India                                   | 297    |
| 5.     | developing countries      | 7      | electronic information resources        | 284    | electronic information resources        | 284    |
| 6.     | online catalogues         | 6      | information & communication technologies| 266    | digital libraries                       | 266    |
| 7.     | adult education workshops | 6      | digital libraries                       | 266    | information & communication technologies| 266    |
| 8.     | scholarly publishing      | 6      | library science                         | 259    | library science                         | 265    |
| 9.     | library information networks | 6   | open access publishing                  | 255    | open access publishing                  | 255    |
| 10.    | agricultural literature   | 5      | information technology                  | 242    | information science                     | 244    |
| 11.    | studies                   | 5      | knowledge management                    | 239    | information technology                  | 244    |
| 12.    | library cooperation       | 5      | electronic journals                     | 235    | knowledge management                    | 239    |
| 13.    | periodicals               | 5      | information science                     | 235    | electronic journals                     | 235    |
| 14.    | library science           | 5      | citation analysis                       | 234    | citation analysis                       | 234    |
| 15.    | chemistry                 | 5      | information retrieval                   | 224    | internet                                | 230    |

**Structure of Institutional collaboration network in LIS research in Indian Universities**

Social Network Analysis (SNA) involves the measurement of particular structural metrics in order to understand the fundamental concepts of social relations[27, 28, 29, 30]. Metrics are used to describe and investigate the connections of a social network[31]. Some of these metrics represent the characteristics of individual nodes whereas others infer a pattern that belongs to the network as a whole[32]. The network dynamics of the collaborating entities are analyzed to map the overall structure (whole network, Figure 2) and also in the individual level with various centrality measures [33,34].

**Network Size:** Among the list of 113 universities with publications, 98 have collaboration with other entities. Rest other universities either do not collaborate or collaborate within their departmental colleagues. So, this study is based on 3,160 records from 98 universities. After removing the duplicate (because of collaborative articles) 2,783 records are finally selected for this study. Among the total 2,127 articles, 1,262 are the collaborative articles. The network is one mode network because we have given equal weightage to the collaborating entities. This whole network consists of are 962 nodes and 2124 ties among the actors. The average degree and the weighted average degree of the nodes in the network show how closely and firmly the actors are tightened together in a network. It is generally observed that in a network, the higher the average degree, the tighter the network is. The whole network in this study has an average degree of 2.208 and the average weighted degree is 3.368. In comparison to the other studies (for example Newman2010, p 237)[35] the cooperation in this network member is not very dense.

**Diameter** of a graph is the largest geodesic distance between any two nodes. In this collaboration network the network diameter is 12 and average path length is 4.973. The diameter shows the span of the network and the average path length shows the minimum distance to travel from one actor to another. The result shows the Indian LIS collaboration network is not very big in comparison to other co-authorship networks in different subjects[35].

The network has 93 connected components and the largest component size consists of 962 actors. All major and productive institutes are in this component. For example, University of Mysore, Bangalore University, DRTC Indian Statistical Institute Bangalore, University of Madras, NISCAIR, NISTADS, IGNOU, Kuvempu University, University of Delhi, Karnataka University are among the prominent actors in this component. The second largest component consists of 17 actors and the most prominent is the Rashtrasant Tukadoji Maharaj Nagpur University. The third largest component has a size of 9 actors and University of Kashmir is the most prominent among them.

**Density** of a collaboration network is simply the proportion of all possible ties that are actually present. For a valued network, density is defined as the sum of the ties divided by the number of possible ties[29]. The density of Indian LIS researchers' network collaboration matrix is .002. That is 0.2 percent of all the possible ties are present. It shows that there is a low level of collaborations among institutions in the Indian LIS research network.

**Clustering Coefficient is** the fraction of paths length between two nodes in the network that are closed[35]. The clustering coefficient (c) 1 implies a perfect transitivity. It happens when a network's components formed cliques and C = 0 infers no closed triads. In this collaboration network average clustering coefficient is 0.149. It means in this collaboration network, collaborators of any one node have very low probability of collaboration with one another. This clustering is much lower than the network cooperation of Medline: 0.066; Los Alamos e-Print Archive: 0.380; SPIRES: 0.726; NCSTRL: 0.496 and so on[35,36] and research collaboration in COLNET Network 0.643[34].

**Figure 4 Collaboration networks of universities in LIS research**

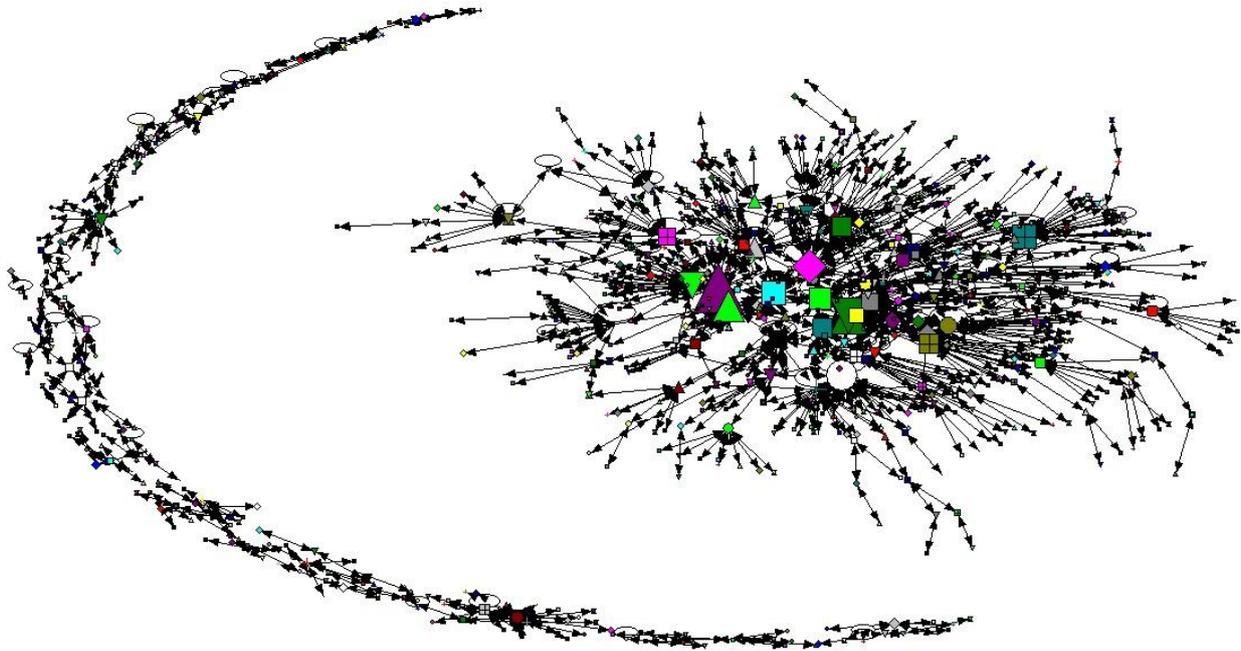

**Micro level structure of the network**

In micro level measurement of a network, Centrality is the important parameter[37, 38, 39]. It signifies the position of a node within a network. Degree Centrality is the most important and simple centrality measure in a network. It denotes the number of edges directly connected to a node. In Indian university LIS research collaboration network of universities are presented in figure 4. The top 10 institutes and their different degree centrality measures are presented in table 4. The highest degree is of University of Mysore (54) followed by University of Delhi (38) and DRTC Indian Statistical Institute, Bangalore (33). The degree centrality of the prominent institutes is shown in table 4. These institutions with the highest degree centrality are more central to the structure of the network. So, these institutions should have a better capacity to influence the others in the network.

Betweenness centrality is a measure for the ability of a node to control the flow of information[40, 41]. Table 4 shows the top institutes with higher betweenness. These institutes with higher betweenness centrality indicate that these are the important actors in the whole network because these institutions are falling on the geodesic paths between other pairs of institutions and have high influence in the collaboration network.

Closeness centrality is the measure the closeness of a node is to all other node in a network. It shows the efficiency of an actor to reach any other actor in a network. Moreover, higher value of this measure means the strength of both the direct and indirect connections of a node with other nodes. The closeness centrality measures are shown in the table 4. As mentioned earlier, closeness centrality measures the distance of an actor to all others in the network, the higher value of these institutes means that these institutes are closer to others. So, they are the more favored institutions in collaboration network.

Eigenvector Centrality is another important centrality measure. It shows the connection of a node to other import node network. In this network University of Mysore has the higher eigenvalue followed by University of Delhi. These institutes are top 10 institutes with higher degree has higher eigenvalue because these institutes itself is important in collaboration and they collaborate among themselves (Table 4)

**Table 3 Different centrality measure of universities**

| Institute | Degree | Id | Betweenness | Id | Closeness | Id | Eigenvector |
|---|---|---|---|---|---|---|---|
| University of Mysore | 54 | University of Delhi | 62593.53 | INFLIB centre | 924482 | University of Mysore | 0.427 |
| University of Delhi | 38 | DRTC | 59916.87 | Jnana Sahyadri, Shanakaraghatta | 924482 | University of Delhi | 0.322 |
| DRTC | 33 | University of Mysore | 45896.43 | Mody institute of management & research | 924482 | Bangalore university | 0.263 |
| University of Kashmir | 32 | Aligarh Muslim University | 36239.03 | Indian Cardamom Research Institute | 924482 | DRTC | 0.259 |
| Karnatak university | 32 | NISTADS | 34353.16 | Integral university | 924482 | NISTADS | 0.211 |
| Sambalpur university | 26 | Bangalore university | 30534.22 | Kurukshetra University | 924482 | Aligarh Muslim University | 0.189 |
| University of madras | 26 | IGNOU | 29628.86 | Sardar Patel university | 924482 | IGNOU | 0.175 |
| Bangalore university | 25 | University of Madras | 27719.91 | Savitribai Phule Pune University | 924482 | DRDO | 0.152 |
| Panjab university | 24 | Indian institute of technology, Kharagpur | 22637.66 | Anand Niketan college | 923521 | NISCAIR | 0.133 |
| Mangalore University | 23 | Karnatak university | 20722.67 | RTM Nagpur university | 923521 | Karnatak university | 0.13 |

**Discussion**

Library and Information Science (LIS) education and research in India is quite old. LIS education and research in India is now conducted mainly in the universities and in two national institute of importance (INSDOC or NISCAIR & DRTC). This article is an attempt to map LIS research in Indian universities. Globally available citation databases (Scopus and Web of Science) have very limited coverage of Indian LIS literature. So, to get the wiser coverage of Indian LIS literature, the publications generated form Library and Science departments of 114 selected Indian universities where extracted from *Library, Information Science & Technology Abstracts* (LISTA). The records were analyzed using Scientometrics, Keyword clustering and Social Network Analysis (SNA) tools.

The research performances of 114 Indian universities and colleges are mapped where postgraduate and PhD level LIS education and trainings are conducted. Among these 114 universities, autonomous colleges and two institutes of national importance, 93 universities have some publications and 21 universities or institutes do not have any publications listed in LISTA. This shows that these institutes are focusing only on teaching. As reflected from LISTA data no research in LIS is being carried out in these universities. Although the publications by Indian LIS researchers started as early as 1964, the contributions of Indian universities were very minimal. During the initial years the two institutions of national importance was predominant in LIS publications. Although the Indian universities published in LIS field, the publication trends were sporadic and inconsistent. The period after the 2000, there was a major boot in LIS research in India. Since the last couple of years about 150 publications are coming up from Indian universities in LIS. Although, the overall LIS publication outputs from universities are growing, the individual university level publication the LIS research, it shows quite low performance. Only the four institutions, the two institutes of national importance (DRTC, NISCAIR) and three universities viz University of Delhi (282), University of Mysore (139) and University of Kashmir (105) have more than 100 publications during 1964-2018. The result shows that, unlike the other subject, research in LIS in universities are perhaps not storng and need to be strengthened.

Articles publish in quality journals which are international in scope are perhaps an possible indicator of research ouput in a field. SRELS Journal of Information Management published by DRTC, Bangalore is the journal where the maximum Indian research outputs are published. This study observed that Indian LIS researchers prefer to publish in Indian journals. Indian LIS journals have many limitations for example many of them are not published in time, inadequate review policy, poor subject coverage, and so on[42]. Hence, the quality of articles published in Indian journals perhaps is not of international standard.

The institutional collaboration network consists of 962 nodes and 2,124 ties among the actors. The network has 93 connected components and the largest component size consists of 427 actors all major and productive institutes are in this component. The productive institutes have formed core group of collaborations among them and this collaboration network may be extended to other universities located in the other locations. It shows that there is a low level of collaborations among institutions in the Indian LIS research network. The network structure also shows that there is a core and periphery structure. Many institutes form collaboration which are not connected with the core institutes.

In micro level measurement of a network, Centrality is the important parameter. It signifies the position of a node within a network. Degree Centrality is the most important and simple centrality measure in a network. It denotes the number of edges directly connected to a node. In Indian university LIS research collaboration network of universities, the highest degree is of University of Mysore, followed by University of Delhi and DRTC Indian Statistical Institute, Bangalore. These institutions with the highest degree centrality are more central to the structure of the network. So, these institutions should have a better capacity to influence the others in the network. University of Delhi, DRTC Indian Statistical Institute Bangalore, University of Mysore and other institutes with higher betweenness centrality indicates that these are the important actors in the whole network because these institutions are falling on the geodesic paths between other pairs of institutions. Among the institutes of national importance DRTC have greater influence than NISCAIR. These two institutions can perhaps lead other organizations in LIS research.

**Concluding Remarks**

The study observed that LIS research programs in Indian universities need to be strengthened as the active research is restricted to a few core institutes. It is interesting to note that unlike other scientific disciplines, the Indian LIS research output is disseminated largely through Indian LIS journals. The reliance of Indian LIS researchers on home grown journals is encouraging. But at the same time, Indian authors should contribute more papers in international journals. This would not only help in doing research of international standards but also would provide much needed opportunities for collaborations with international institutes.

40. Abbasi A, Hossaina L, and Leydesdorff L, Betweenness centrality as a driver of preferential attachment in the evolution of research collaboration networks, *Journal of Informetrics,* 6 (3) (2012) 403-412.
41. Leydesdorff L, Betweenness Centrality as an Indicator of the Interdisciplinarity of Scientific Journals, *Journal of the American Society for Information Science and Technology, 58*(9) (2007) 1303-1309.
42. Vishwakarma P and Mukherjee B, Developing Qualitative Indicators for Journal Evaluation: Case Study of Library Science Journals of SAARC Countries. *DESIDOC Journal of Library & Information Technology*, 34(2) (2014) 152-161.

20